\documentclass[useAMS,usenatbib]{mn2e}
\usepackage{epsfig}

\title[Simulating seed BH formation: influence of X-rays]{Simulating the formation of massive seed black holes in the early Universe. III: The influence of X-rays}
\author[S.~C.~O. Glover]{Simon C. O. Glover \\
Universit\"{a}t Heidelberg, Zentrum f\"{u}r Astronomie, Institut f\"{u}r Theoretische Astrophysik, \\ Albert-Ueberle-Stra{\ss}e 2, 
69120 Heidelberg, Germany
}

\begin{document}
\maketitle

\begin{abstract}
The direct collapse black hole (DCBH) model attempts to explain the observed number density of supermassive black holes
in the early Universe by positing that they grew from seed black holes with masses of $10^{4}$--$10^{5} \: {\rm M_{\odot}}$
that formed by the quasi-isothermal collapse of gas in metal-free protogalaxies cooled by atomic hydrogen emission.
For this model to work, H$_{2}$ formation must be suppressed in at least some of these systems by a strong 
extragalactic radiation field. The predicted number density of DCBH seeds is highly sensitive to the minimum value of 
the ultraviolet (UV) flux required to suppress H$_{2}$ formation, $J_{\rm crit}$. In this paper, we examine how the value of
$J_{\rm crit}$ varies as we vary the strength of a hypothetical high-redshift X-ray background. We confirm earlier findings
that when the X-ray flux $J_{\rm X}$ is large, the critical UV flux scales as $J_{\rm crit} \propto J_{\rm X}^{1/2}$. We also
carefully explore possible sources of uncertainty arising from how the X-rays are modelled. We use a reaction-based reduction
technique to analyze the chemistry of H$_{2}$ in the X-ray illuminated gas and identify a critical subset of 35 chemical 
reactions that must be included in our chemical model in order to predict $J_{\rm crit}$ accurately. We further show that
$J_{\rm crit}$ is insensitive to the details of how secondary ionization or He$^{+}$ recombination are modelled,
but does depend strongly on the assumptions made regarding the column density of the collapsing gas.
\end{abstract}


\section{Introduction}
The presence of supermassive black holes (SMBHs) with masses $\sim 10^{9} \: {\rm M_{\odot}}$ and above
at redshifts $z > 6$ \citep[see e.g.][]{mort11,wu15,venemans15} is challenging to explain within the context of 
the standard $\Lambda$CDM cosmological model. The problem is one of timescales: if the SMBHs form from 
seed black holes with masses $M \sim 10 \: {\rm M_{\odot}}$, produced as remnants of the earliest generations of 
massive stars, which then grow by accretion, then there does not seem to be enough time for them to grow to 
the observed masses by $z \sim 6$ unless one invokes a length period of super-Eddington accretion 
\citep[see e.g.\ the recent review by][and references therein]{jh16}.

Because of this, an alternative model -- the direct collapse black hole (DCBH) model -- has recently been
attracting significant attention. In this model, the seeds of the high redshift SMBHs are not stellar mass black 
holes, but instead are intermediate mass black holes (IMBHs) with masses of $10^{4}$--$10^{5} \: {\rm M_{\odot}}$
that form in some of the earliest protogalaxies. The basic idea is simple. If gas in a protogalaxy with virial
temperature $T_{\rm vir} > 10^{4} \: {\rm K}$ -- an ``atomic cooling halo'' -- is illuminated by a strong external radiation field, then the formation
of molecular hydrogen (H$_{2}$) within the protogalaxy can be almost completely suppressed \citep{om01,bl03}.
If the gas is also metal-free, i.e.\ if it has not been enriched by metals from an earlier generation of stars, then the 
end result is that the gas is unable to cool much below the temperature ($T \sim 5000$--$6000$~K) at which
Lyman-$\alpha$ and H$^{-}$ cooling become ineffective \citep{om01,osh08,ssg10}. It therefore collapses quasi-isothermally,
eventually forming a protostar surrounded by a massive accretion disc \citep{iot14,bec15,lsh16}. The high temperature and
correspondingly high Jeans mass inhibit fragmentation and also lead to gas accreting onto the newly formed
protostar at an extremely high rate, $\dot{M} \sim 1 \: {\rm M_{\odot} \: yr^{-1}}$. The outcome of this process is not entirely certain, 
but the most likely possibilities are either the formation of a supermassive star with $M \sim 10^{5} \: {\rm M_{\odot}}$
that collapses after $10^{5}$--$10^{6} \: {\rm yr}$ to form a massive black hole \citep{hos13}, or the formation of a quasi-star,
an optically thick massive core whose centre has already collapsed to form a black hole \citep{begel06,begel10,schleicher13}. 
In either case, the end result is an IMBH that subsequently grows into a high-redshift SMBH. 

The much larger mass of the seed black hole in the DCBH model greatly alleviates the timing problem
referred to above, since considerably less time is required to grow a $10^{4}$--$10^{5} \: {\rm M_{\odot}}$ seed
to $\sim 10^{9} \: {\rm M_{\odot}}$ than to grow a $10 \: {\rm M_{\odot}}$ seed to the same final mass. 
However, one major uncertainty in this model is the strength of the radiation field that is required in order to completely
suppress H$_{2}$ formation and prevent fragmentation of the gas. This is often parameterized in terms of $J_{21}$, 
the specific intensity of the radiation field at the Lyman limit in units of $10^{-21} \: {\rm erg \, s^{-1} \, cm^{-2} \, Hz^{-1}
\, sr^{-1}}$, with the minimum value of $J_{21}$ required for direct collapse being written as $J_{\rm crit}$. The value
of $J_{\rm crit}$ is important as it strongly influences the likelihood of finding a suitable site for DCBH formation within
any given cosmological volume. For instance, \citet{dfm14} find that varying the value of $J_{\rm crit}$ by only a factor
of a few changes the predicted number density of DCBH seeds in their model by orders of magnitude. More recently,
\citet{it15} have argued that the probability of finding a suitable halo scales as $J_{\rm crit}^{-5}$. Accurate determination
of the value of $J_{\rm crit}$ is therefore important for assessing the viability of the DCBH model: if the predicted
comoving number density of DCBH seeds is too small (i.e.\ smaller than the observed comoving number density of
SMBHs at $z \sim 6$), then the model will be unable to explain the observed high-redshift SMBHs.\footnote{If the predicted 
number density of DCBH seeds is far too large, then this may also be problematic, but is not immediately fatal for the model, 
since there is no guarantee that all of the seeds will accrete sufficient gas to become SMBHs by $z \sim 6$.}

Values quoted in the literature for $J_{\rm crit}$ span a wide range, from $\sim 20$ \citep{io11,Glover15a} to $10^{5}$ 
\citep{om01}. There are several reasons for this large scatter. Firstly, the value of $J_{\rm crit}$ is known to depend 
strongly on the spectral shape of the extragalactic radiation field \citep{sbh10,soi14,latif15,ak15,ag16,ag16b}.
In part, this is because this quantity is a relatively crude way of specifying the quantities that actually matter physically, 
the H$^{-}$ photodetachment rate and the H$_{2}$ photodissociation rate; if we change the shape of the extragalactic 
radiation field, then generally we will also need to adjust the value of $J_{21}$ in order to recover the same values for 
these rates. 

Secondly, the value of $J_{\rm crit}$ is sensitive to the way in which the effects of H$_{2}$ self-shielding are treated. 
Typically, this is modeled using a simple shielding function that parameterises the results of more detailed calculations.
However, the two functions most commonly used, introduced by \citet{db96} and \citet{wgh11}, respectively, yield 
values for $J_{\rm crit}$ that can differ by up to an order of magnitude \citep{soi14}. In addition, \citet{hartwig15} have
shown that properly accounting for the velocity structure of the gas in the protogalaxy reduces the effectiveness of
self-shielding and decreases $J_{\rm crit}$ by roughly a factor of two. 

Thirdly, the value of $J_{\rm crit}$ that we derive from numerical models of direct collapse can be sensitive to details
of the chemical modelling: the reactions included in the chemical network and the choice of reaction rate coefficients
\citep{Glover15a,Glover15b}.

Finally, several studies have shown that $J_{\rm crit}$ can also depend on the strength of the X-ray background radiation
field illuminating the protogalaxy \citep{io11,it15,latif15}. X-rays penetrate easily into high column density gas and hence
are able to increase the fractional ionization of the gas throughout the protogalaxy. The additional free electrons catalyze
H$_{2}$ formation via the H$^{-}$ pathway
\begin{eqnarray}
{\rm H + e^{-}} & \rightarrow & {\rm H^{-} + \gamma}, \\
{\rm H + H^{-}} & \rightarrow & {\rm H_{2} + e^{-}},
\end{eqnarray}
and hence a larger number of LW photons are required in order to completely suppress H$_{2}$ cooling. We therefore
expect that increasing the strength of the X-ray background should increase $J_{\rm crit}$. The studies cited above
confirm this expectation, but do not agree when it comes to the strength of the effect. 

In this paper we re-examine the impact of X-rays on DCBH formation using a series of simple one-zone models.
Building on the work in \citet{Glover15a}, we identify the subset of chemical reactions that it is necessary to include
in the chemical network used to model DCBH formation in order to derive reliable results for $J_{\rm crit}$. 
We also examine how sensitive $J_{\rm crit}$ is to the choices made when constructing a model for the effects of
X-ray photoionization (e.g.\ the choice of photoionization cross-sections, the treatment of secondary ionization, or
the assumptions regarding shielding), and demonstrate that some of these choices can have a substantial impact 
on the values of $J_{\rm crit}$ we derive when the X-ray background is strong.

The structure of the paper is as follows. In Section~\ref{numerics}, we present the numerical method and initial
conditions used for our one-zone calculations. We also discuss the reaction-based reduction technique that we
use to identify the subset of essential chemical reactions. In Section~\ref{sec:Jc}, we explore how $J_{\rm crit}$
varies as a function of the X-ray flux $J_{\rm X}$ in our fiducial model. In Section~\ref{sec:reduce}, we apply
our reaction-based reduction technique to the results of our fiducial model and discuss the resulting ``reduced''
chemical network that this allows us to construct. In Section~\ref{sec:uncertain}, we examine a number of
possible sources of uncertainty in our results, including the treatment of X-ray shielding and the spectral
properties of the incident X-ray background. We compare our results with those from previous studies in 
Section~\ref{sec:prev} and conclude with a brief summary in Section~\ref{sec:end}.

\section{Numerical method}
\label{numerics}

\subsection{One-zone model}
We model the thermal and chemical evolution of the gravitationally-collapsing primordial gas in our simulations using an updated
version of the one-zone model presented in \citet{Glover15a}, which itself was derived from a model originally developed by
\citet{Glover09}. Full details of the model are given in these two papers, and so here we simply remind the reader of the basic
features and describe the changes that we have made in order to account for the effects of an X-ray background.

The gas density in the model is assumed to evolve according to 
\begin{equation}
\frac{{\rm d}\rho}{{\rm d}t} = \frac{\rho}{t_{\rm ff}},
\end{equation}
where $t_{\rm ff} = (3\pi / 32 G \rho)^{1/2}$ is the free-fall time of the gas. The internal energy density $e$ evolves as 
\begin{equation}
\frac{{\rm d}e}{{\rm d}t} = \frac{p}{\rho^{2}} \frac{{\rm d}\rho}{{\rm d}t} + \Gamma - \Lambda,
\end{equation}
where $\Gamma$ and $\Lambda$ are the radiative heating and cooling rates per unit volume, respectively, and
$p$ is the gas pressure. $\Gamma$ and $\Lambda$ are computed using the detailed atomic and molecular cooling
function described in \citet{Glover09} and updated in \citet{Glover15a}. The only significant change that we have made
in the present study is the inclusion of the effects of X-ray heating, which we discuss in the next section. 

The chemical evolution of the gas is followed using an updated version of the \citet{Glover15a} chemical network.
The only change that we have made to this network compared to the version presented in \citet{Glover15a} is the
inclusion of eight X-ray induced chemical reactions that were not previously accounted for: 
the X-ray photoionization of H, D, He, He$^{+}$, Li, H$_{2}$ and HD, as well as the double photoionization of He
\begin{equation}
{\rm He + \gamma}  \rightarrow  {\rm He^{++} + e^{-} + e^{-}}.
\end{equation}
We describe how we compute the rates of these reactions in Section~\ref{sec:xray} below. Our complete network
therefore consists of 30 different primordial chemical species linked by a total of 400 reactions. 

We perform simulations with two different models for the spectral shape of the radiation field at energies
$h\nu \leq 13.6$~eV: a $10^{4}$~K diluted black-body spectrum (hereafter a T4 spectrum) and a 
$10^{5}$~K diluted black-body spectrum (hereafter a T5 spectrum). Following \citet{har00}, we 
normalize these spectra by specifying their mean specific intensity at the Lyman limit in units of 
$10^{-21} \: {\rm erg s^{-1} cm^{-2} Hz^{-1} sr^{-1}}$, a quantity that is commonly written as $J_{21}$.
For both spectra, we assume that there is zero flux at energies above 13.6~eV and below the minimum
X-ray energy considered (see Section~\ref{sec:xrspec} below), owing to efficient absorption by atomic hydrogen in
the intergalactic medium. Although both of these spectral shapes are highly idealized compared to 
the spectral energy distributions produced by more realistic models of high redshift galaxies \citep{soi14,ak15,ag16,ag16b},
they nevertheless are useful for the kind of exploratory study presented in this paper, because they are
simple and they allow us to probe the two different physics regimes relevant for the radiative suppression
of H$_{2}$ cooling in the DCBH scenario. In the case of the T5 spectrum, H$_{2}$ photodissociation is
the dominant effect and in our fiducial model with this spectrum, H$_{2}$ cooling is totally suppressed
provided that $J_{21} \ge J_{\rm crit} \simeq 1630$ \citep{Glover15a}, corresponding to an H$_{2}$ 
photodissociation rate in unshielded gas $k_{\rm dis} \simeq 2.1 \times 10^{-9} \: {\rm s^{-1}}$. With the T4 spectrum,
on the other hand, complete suppression of H$_{2}$ cooling occurs when $J_{21} > 18$, corresponding
to an H$_{2}$ photodissociation rate of only $4.9 \times 10^{-11} \: {\rm s^{-1}}$. The reason for this is that in this case,
H$^{-}$ photo-detachment by lower energy photons, which occurs at a rate $k_{\rm pd} = 3.6 \times 10^{-6} \: {\rm s^{-1}}$, 
dramatically reduces the H$_{2}$ formation rate, meaning that a much smaller H$_{2}$ photodissociation 
rate is required to reduce the H$_{2}$ abundance to the same level. Physically, therefore, changes in 
the value of $J_{\rm crit}$ brought about by X-ray photoionization actually correspond to changes in the
values of $k_{\rm dis}$ and/or $k_{\rm pd}$ that are required in order to completely suppress H$_{2}$
cooling. 

We account for the effects of H$_{2}$ self-shielding using the \citet{wgh11} shielding function. In order to do so, 
we need to estimate the H$_{2}$ column density. We do this using the same approach as in \citet{Glover15a,Glover15b}: 
we assume that the dominant contribution comes from a constant density core with a radius equal to the Jeans length, 
$\lambda_{\rm J}$, so that $N_{\rm H_{2}} = n_{\rm H_{2}} \lambda_{\rm J}$. We do not account for any other
sources of opacity, as we expect the continuum opacity of primordial gas to be extremely small at the densities
and temperatures relevant for our study \citep{md05}.

\subsection{Treatment of X-rays}
\label{sec:xray}
\subsubsection{X-ray photoionization}
The photoionization rate of a species $m$ due to direct X-ray photoionization can be written in the form
\begin{equation}
R_{{\rm d}, m} = \int_{\nu_{\rm min}}^{\nu_{\rm max}} \frac{4\pi J_{\rm X}(\nu)}{h\nu} \sigma_{m}(\nu) e^{-\tau_{\nu}} {\rm d}\nu,
\end{equation}
where $\nu_{\rm min}$ and $\nu_{\rm max}$ are the minimum and maximum X-ray frequencies considered, 
$J_{\rm X}$ is the mean specific intensity of the X-ray background (discussed in more detail in Section~\ref{sec:xrspec}
below), $\sigma_{m}(\nu)$ is the photoionization
cross-section of species $m$ at frequency $\nu$ and $\tau_{\nu}$ is the optical depth of the gas at the same frequency.
In principle, many different processes contribute towards $\tau_{\nu}$, but in practice, in the conditions relevant
for DCBH formation, the dominant contributions come from the photoionization of H and He, so that
\begin{equation}
\tau_{\nu} = \sigma_{\rm H}(\nu) N_{\rm H} + \sigma_{\rm He}(\nu) N_{\rm He},
\end{equation}
where $\sigma_{\rm H}$ and $\sigma_{\rm He}$ are the photoionization cross-sections of H and He, and
$N_{\rm H}$ and $N_{\rm He}$ are corresponding column densities. In our fiducial model, we adopt
expressions for the photoionization cross-sections taken from \citet{ost89} for H and He$^{+}$ and \citet{ysd98,ysd01} 
for He (both single and double photoionization) and H$_{2}$. We assume that the photoionization cross-section
of D is the same as that for H, and that the value for HD is the same as that for H$_{2}$. Finally, for Li, we use
the cross-section given in \citet{vf96}.

In Section~\ref{sec:cross}, we explore the impact of using the H and He photoionization cross-sections given in 
\citet{vf96} in place of the \citet{ost89} and \citet{ysd98,ysd01} values,  since the \citet{vf96} values have been used in
a number of previous studies \citep[e.g.][]{latif15}. 

To compute $N_{\rm H}$ and $N_{\rm He}$, we adopt the same approach as for H$_{2}$ above, and set
$N_{\rm H} = n_{\rm H} \lambda_{\rm J}$ and $N_{\rm He} = n_{\rm He} \lambda_{\rm J}$, where 
$\lambda_{\rm J}$ is the local value of the Jeans length.

In addition, it is also necessary to account for secondary ionizations caused by the energetic photoelectrons
produced by direct X-ray photoionization.  The secondary ionization rate of a species $m$ can be written as
\begin{equation}
R_{{\rm s}, m} = \sum_{i} \frac{n_{i}}{n_{m}} R_{{\rm d}, i} \bar{N}_{{\rm ion}, m, i}
\end{equation}
where $i = {\rm H, D, He, He^{+}, Li}, {\rm H_{2}}$ or HD, $R_{{\rm d}, i}$ is the direct photoionization rate of species
$i$, and $\bar{N}_{{\rm ion}, m, i}$ is the number of secondary ionizations of species $m$ per primary ionization
of species $i$. To compute $\bar{N}_{{\rm ion}, m, i}$, we first make the simplifying assumption that secondary
ionization of any species other than H or He is negligible. This is valid provided that the fractional abundances
of He$^{+}$ and H$_{2}$ are much smaller than those of He and H, which is always true for gas in the regime
of interest for this study. We then have
\begin{equation}
\bar{N}_{{\rm ion, H}, i} = R_{{\rm d}, i}^{-1} \int_{\nu_{\rm min}}^{\nu_{\rm max}} \frac{4\pi J_{\rm X}}{h\nu} 
\sigma_{i} e^{-\tau_{\nu}} N_{\rm ion, H}(E, x) \,
{\rm d}\nu,
\end{equation}
where $N_{\rm ion, H}(E, x)$ is the mean number of ionizations produced by an electron with
energy $E = h\nu - h\nu_{i}$ in gas with fractional ionization $x$, and $h\nu_{i}$ is the ionization threshold
for species $i$. Similarly, for helium we have
\begin{equation}
\bar{N}_{{\rm ion, He}, i} = R_{{\rm d}, i}^{-1} \int_{\nu_{\rm min}}^{\nu_{\rm max}} \frac{4\pi J_{\rm X}}{h\nu} 
\sigma_{i} e^{-\tau_{\nu}} N_{\rm ion, He}(E, x) \,
{\rm d}\nu
\end{equation}
In our fiducial model, we compute $N_{\rm ion, H}(E, x)$ and $N_{\rm ion, He}(E, x)$
using the values tabulated by \citet{dyl99} for a mixed gas of H and He. However, in Section~\ref{sec:second} 
we examine how much our results change if we instead us the simpler treatment given in \citet{ss85}.

\subsubsection{X-ray heating}
The heating rate of the gas due to the photoelectrons produced by X-ray photoionization of a species $m$
can be written as 
\begin{equation}
H_{m} = \int_{\nu_{\rm min}}^{\nu_{\rm max}} \frac{4\pi J_{\rm X}(\nu)}{h\nu} \sigma_{m}(\nu) e^{-\tau_{\nu}} 
E \, f_{\rm heat}(E, x) \, {\rm d}\nu,
\end{equation}
where $E = h\nu - h\nu_{m}$ is the energy of the photoelectron and $f_{\rm heat}(E, x)$ is the fraction of this
energy that is deposited as heat. In our fiducial model, we compute this using the values tabulated
by \citet{dyl99} for a mix of H and He, but we also explore the effects of using instead the prescription
given in \citet{ss85}. Finally, given the heating rate for each species, the total X-ray heating rate per
unit volume then follows as
\begin{equation}
\Gamma_{\rm X} = \sum_m H_{m} n_{m},
\end{equation}
where $m = {\rm H, D, He, He^{+}, Li}, {\rm H_{2}}$ or HD. In practice, we find that the dominant
contributions generally come from H and He photoionization, since at the densities of interest these
species are far more abundant than any of the others.

\subsubsection{X-ray spectrum}
\label{sec:xrspec}
We assume for simplicity that the mean specific intensity of the X-ray background, $J_{\rm X}$, can
be represented as a power-law:
\begin{equation}
J_{\rm X} = J_{\rm X, 21} \times 10^{-21} \left(\frac{\nu}{\nu_{0}}\right)^{-\alpha} \, {\rm erg \, s^{-1} \, cm^{-2} \, sr^{-1} \, Hz^{-1}},
\end{equation}
where $h\nu_{0} = 1 \: {\rm keV}$. In our fiducial model, we set $\alpha = 1.5$, following \citet{gb03}, but in 
Section~\ref{res:xray} we study the effects of varying it between $\alpha = 1.0$ (as in \citealt{km05} and \citealt{jeon14}) and 
$\alpha = 1.8$ (the value adopted by \citealt{it15}, based on \citealt{swartz04}). 

We assume that absorption close to the X-ray sources or in the intergalactic medium attenuates all X-rays below a 
minimum energy cutoff which by default we take to be $h\nu_{\rm min} = 0.5$~keV. In Section~\ref{res:xray}, we explore 
the effects of varying this default value. In all of our models, we set the maximum photon energy
to $h\nu_{\rm max} = 30$~keV, but we note that as photons with energies close to $h\nu_{\rm max}$ make a 
negligible contribution to the photoionization and X-ray heating rates, our results are insensitive to this value.

\subsubsection{Strength of the X-ray background}
The strength and spatial variation of the X-ray background at high redshift are poorly constrained. In the local
Universe, high mass X-ray binaries dominate the X-ray emission from star-forming galaxies, and so there is
a good correlation between the X-ray luminosity and the star formation rate \citep[see e.g.][]{gb03,ggs03,min14}.
If we assume that the same correlation holds for the earliest star-forming galaxies, that these galaxies are
forming stars at a roughly constant rate with a standard Salpeter initial mass function, and that the escape
fraction of the Lyman-Werner photons produced by these galaxies is close to 100\%, then one can show 
that $J_{\rm X, 21} \sim 4 \times 10^{-6} J_{21}$ \citep{it15}, although this value can change by up to 50\%
depending upon what one assumes for the shape of the UV and X-ray spectra.  For the range of UV field
strengths considered in this study, this expression yields values of $J_{\rm X, 21}$ in the range 
$10^{-4}$--$0.1$. However, it is possible that the number of high mass X-ray binaries per unit of star 
formation is much higher at high redshift \citep{f13a,f13b}, implying that the ratio of $J_{\rm X, 21}$ to
$J_{21}$ could easily be an order of magnitude or more higher. Therefore, in this study we adopt values
of $J_{\rm X, 21}$ spanning the range from  $10^{-4}$  to 1.

\subsection{Initial conditions}
As in \citet{Glover15a}, we take the initial gas density to be $n_{0} = 0.3 \: {\rm cm^{-3}}$, a value comparable
to the mean density of a virialized protogalaxy at $z = 20$. In our previous study, we examined several different choices
for the initial temperature and chemical composition of the gas, but found that they all yielded very similar results for
$J_{\rm crit}$. In this paper, we therefore consider for simplicity only a single set of values. Our choices are the same
as those used in runs 2 and 5 in \citet{Glover15a}: an initial temperature $T_{0} = 8000 \: {\rm K}$, an initial free
electron abundance $x_{\rm e, 0} = 2 \times 10^{-4}$ and an inital H$_{2}$ fractional abundance $x_{\rm H_{2}, 0}
= 2 \times 10^{-6}$. For deuterium and lithium, we adopt elemental abundances relative to hydrogen given by
$A_{\rm D} = 2.6 \times 10^{-5}$ and $A_{\rm Li} = 4.3 \times 10^{-10}$, respectively
\citep{cyburt04}. We assume that the gas is electrically neutral and that the initial D$^{+}$ abundance
is a factor of  $A_{\rm D}$ smaller than the initial H$^{+}$ abundance, so that
\begin{equation}
x_{\rm e, 0} = \frac{1}{1 + A_{\rm D}} x_{\rm H^{+}, 0}  + \frac{A_{\rm D}}{1 + A_{\rm D}} x_{\rm D^{+}, 0}.
\end{equation}
Lithium and helium are assumed to start in neutral atomic form. We set the starting value of the HD abundance to
be $x_{\rm HD, 0} = A_{\rm D} x_{\rm H_{2}, 0}$, and initialize the abundances of all of the other chemical species
in our model to zero.

\begin{table}
\caption{List of simulations \label{tab:xray}}
\begin{tabular}{ccc}
\hline
$h\nu_{\rm min}$ &  & \\
(keV) & $\alpha$ & Notes \\
\hline
0.5 & 1.5 & Fiducial model \\
0.1 & 1.5 & \\
1.0 & 1.5 & \\
2.0 & 1.5 & \\
0.5 & 1.0 & \\
0.5 & 1.8 & \\
0.5 & 1.5 & $N_{\rm H} = n_{\rm H} \lambda_{\rm J} / 2$ \\
0.5 & 1.5 & Case A He$^{+}$ recombination \\
0.5 & 1.5 & Case B He$^{+}$ recombination \\
0.5 & 1.5 & Secondary ion.\ from SS85 \\
0.5 & 1.5 & $\sigma_{\rm H}$ and $\sigma_{\rm He}$ from V96 \\
\hline
\end{tabular}
\\ SS85 = \citet{ss85}
\\ V96 = \citet{vf96}
\end{table}

With the initial chemical composition, density and temperature fixed, the remaining free parameters in our one-zone
model are those controlling the strength and shape of the optical/UV spectrum ($J_{21}$, choice of T4 or T5 spectrum)
and the X-ray spectrum ($J_{\rm X, 21}$, $h\nu_{\rm min}$, $\alpha$). In our fiducial setup, we set $h\nu_{\rm min} = 1.0 \:
{\rm keV}$ and $\alpha = 1.5$, but we also examine additional combinations of these two parameters, as summarized in Table~\ref{tab:xray}.
We perform runs with both T4 and T5 spectra, and in each case explore a wide range of different
values of $J_{\rm X, 21}$, ranging from $10^{-4}$ to 1. For each combination of optical/UV spectrum, $J_{\rm X, 21}$, $h\nu_{\rm min}$
and $\alpha$, we determine $J_{\rm crit}$ using the method described in Section~\ref{jcrit_det} below.

\subsection{Determination of $J_{\rm crit}$}
\label{jcrit_det}
We determine the value of $J_{\rm crit}$ in a given model using the same 
binary search approach as in \citet{Glover15a}. We first select two values of $J_{21}$, one very small
($J_{\rm 21, low}$) and the other very large ($J_{\rm 21, high}$), 
that we can be confident bound the value of $J_{\rm crit}$ from below and above.
We next calculate a new value of $J_{21}$ using the equation
\begin{equation}
J_{\rm 21, new} = \left(J_{\rm 21, low} \times  J_{\rm 21, high} \right)^{1/2}.
\end{equation}
We run a simulation with this new value of $J_{21}$, following the collapse of the gas until the H nuclei
number density reaches $n = 10^{8} \: {\rm cm^{-3}}$. We then examine the outcome of this simulation. 
If H$_{2}$ cooling is suppressed and the temperature remains $\sim 6000$~K or higher throughout the collapse of the gas,
then we know that $J_{\rm 21, new} \ge J_{\rm crit}$. In this case, we adopt $J_{\rm 21, new}$ as our
new value of $J_{\rm 21, high}$. On the other hand, if H$_{2}$ cooling is not suppressed and the gas
cools to temperatures $T \ll 6000$~K during the collapse, then we know that $J_{\rm 21, new} < J_{\rm crit}$
and hence adopt it as our new value of $J_{\rm 21, low}$. We then calculate a new value of 
$J_{\rm 21, new}$ and repeat the whole procedure. We continue in this fashion until the difference 
between $J_{\rm 21, low}$ and $J_{\rm 21, high}$ is less than 0.2\%. At this point we stop, adopting the
final value of $J_{\rm 21, new}$ as our estimate for $J_{\rm crit}$. 

\subsection{Identifying the most important chemical reactions}
\label{reduct}
The detailed model of primordial chemistry that we use in the calculations presented in this paper is
too large to use directly in 3D hydrodynamical models of DCBH formation without incurring excessive
computational overhead. Therefore, one of the main goals of this paper is to identify an appropriate
``reduced'' chemical network that includes only those reactions and species that are required for
accurately following the time evolution of the H$_{2}$ abundance. In \citet{Glover15a}, we already
carried out an analysis of this kind for the case where $J_{\rm X} = 0$. In the present paper, we merely
extend this analysis to also consider the case where $J_{\rm X} > 0$. Full details of our approach are
given in \citet{Glover15a}, but we summarize the main points here.

We identify important reactions and species using the  reaction-based reduction technique developed by 
\citet{Wiebe03}. Starting from a prescribed set of initial conditions, we first compute the chemical and 
thermal evolution of the gas for a given value of $J_{21}$ using our one-zone model. We then select our
starting set of important species, which in the present case contains only a single member (H$_{2}$).
For each species in our set, we compute the production and destruction processes
for that species at a series of different output times using the chemical abundances and temperature
computed by our one-zone code. We identify ``important'' processes using the reaction-weighting
procedure described in detail in \citet{Wiebe03} and \citet{Glover15a}. If the set of important reactions
includes chemical species that are not in our current set of important species, then we add them to the
set and repeat the analysis, proceeding in this fashion until there are no more species or reactions that 
need to be added. 

This reduction procedure generates for each output time a list of the reactions that are required in
order to accurately model the H$_{2}$ abundance at that time. The composition of this list may
change over time as the density, temperature and chemical abundances change, and so we generate
a final list, valid for the whole range of times modelled in our simulations, by combining the individual
lists.

The only free parameter in this method enters when we decide whether or not to classify a given 
reaction as important. We compare the weight for the reaction with a cut-off value $\epsilon$ and
retain only those reactions with weights $> \epsilon$.  As in \citet{Glover15a}, we conservatively
set $\epsilon = 10^{-4}$. This yields a set of reactions that allow us to determine $J_{\rm crit}$ to 
within an accuracy of around 1\% when compared  to the results of models run with the full chemical 
network.  

\section{The effect of X-rays on $J_{\rm crit}$}
\label{sec:Jc}
\begin{figure}
\includegraphics[width=3.2in]{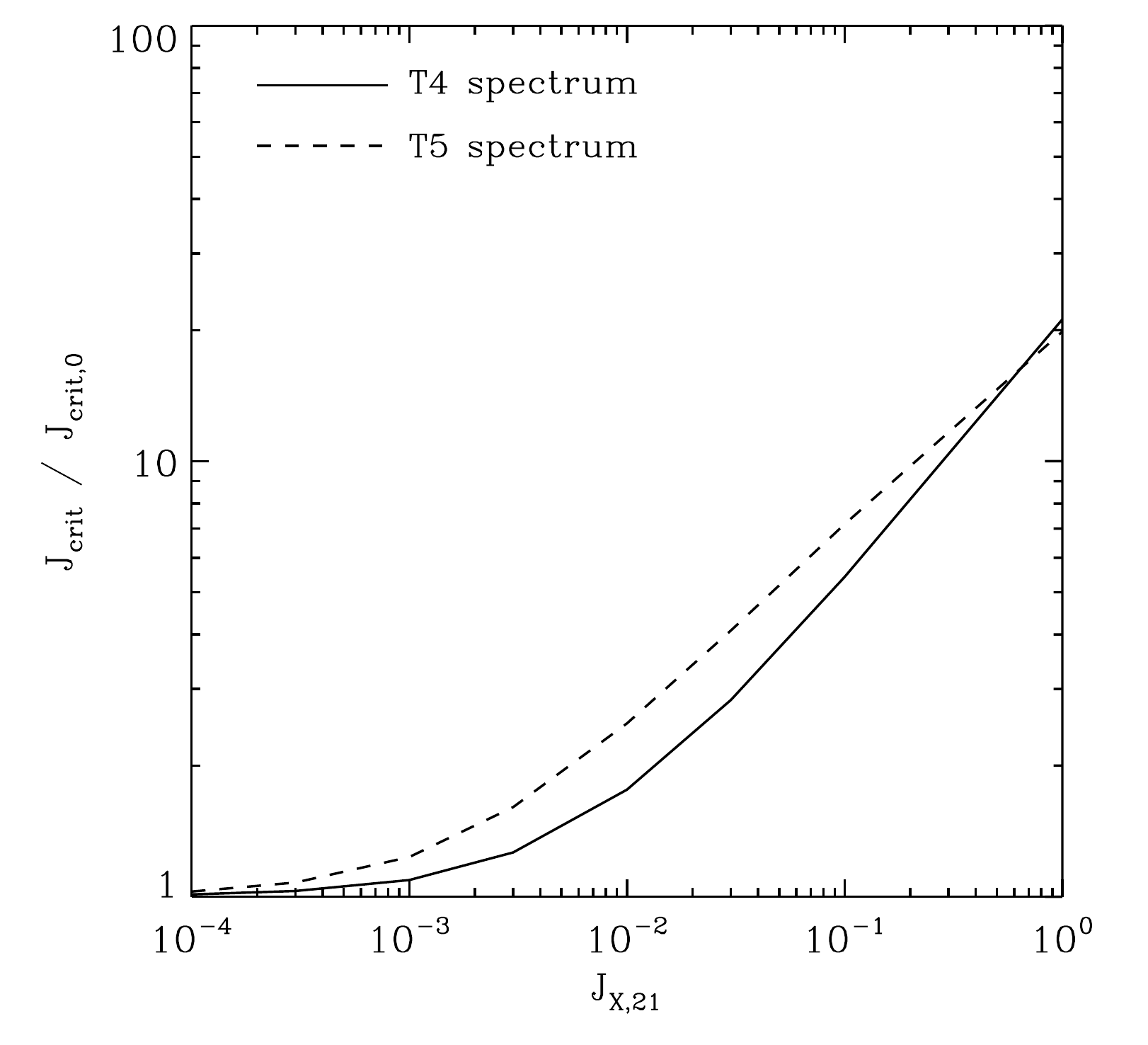}
\caption{Dependence of $J_{\rm crit}$ on $J_{\rm X, 21}$, computed for our fiducial X-ray model
for both a T4 spectrum (solid line) and a T5 spectrum (dashed line). In both cases, we normalize
the values of $J_{\rm crit}$ by the value we obtain in the absence of X-rays, $J_{\rm crit, 0}$.
This value depends strongly on the choice of spectrum: for the T4 spectrum, $J_{\rm crit, 0} = 18.0$,
while for the T5 spectrum, $J_{\rm crit, 0} = 1630$. However, the growth in $J_{\rm crit}/J_{\rm crit, 0}$
with increasing $J_{\rm X, 21}$ is very similar in both cases. \label{Jcrit_JX_def}}
\end{figure}

For our fiducial X-ray model, we explore a range of different values of $J_{\rm X, 21}$, ranging from $10^{-4}$ 
to 1. For each value of $J_{\rm X, 21}$, we determine the corresponding value of $J_{\rm crit}$ using the
method described in Section~\ref{jcrit_det}. Our results are summarized in Table~\ref{tab:jcrit}, while in 
Figure~\ref{Jcrit_JX_def} we show how the
relative increase in $J_{\rm crit}$, i.e.\ the value divided by the value in the absence of X-rays, $J_{\rm crit, 0}$,
evolves as we alter $J_{\rm X, 21}$.  We find,
in agreement with previous studies, that when $J_{\rm X, 21}$ is small, X-rays have very little influence on
$J_{\rm crit}$ \citep{latif15,it15}. However, as we increase the X-ray flux, the value of $J_{\rm crit}$ steadily
increases. The effect is relatively small for $J_{\rm X, 21} < 0.01$, but at higher values $J_{\rm crit} \propto
J_{\rm X, 21}^{1/2}$. We also see that the relative increase in $J_{\rm crit}$ is very similar for the T4 and
T5 spectra, even though the absolute values of $J_{\rm crit}$ differ by almost two orders of magnitude.

\begin{table}
\caption{Dependence of $J_{\rm crit}$ on $J_{\rm X, 21}$ for
our fiducial X-ray model \label{tab:jcrit}}
\begin{tabular}{rccc}
\hline
$J_{\rm X, 21}$ & Spectrum & $J_{\rm crit}$ & $J_{\rm crit} / J_{\rm crit, 0}$ \\
\hline
0.0 & T4 & 18.0 & 1.00 \\
$10^{-4}$ & T4 & 18.5 & 1.03 \\
$3 \times 10^{-4}$ & T4 & 19.4 & 1.08 \\
$10^{-3}$ & T4 & 22.2 & 1.23 \\
$3 \times 10^{-3}$ & T4 & 28.9 & 1.61 \\
0.01 & T4 & 45.0 & 2.50 \\
0.03 & T4 & 73.5 & 4.08 \\
0.1 & T4 & 129 & 7.17 \\
1.0 & T4 & 357 & 19.8 \\
\hline
0.0 & T5 & 1630 & 1.00 \\
$10^{-4}$ & T5 & 1650 & 1.01 \\
$3 \times 10^{-4}$ & T5 & 1680 & 1.03 \\
$10^{-3}$ & T5 & 1780 & 1.09 \\
$3 \times 10^{-3}$ & T5 & 2060 & 1.26 \\
0.01 & T5 & 2870 & 1.76 \\
0.03 & T5 & 4610 & 2.83 \\
0.1 & T5 & 8840 & 5.42 \\
1.0 & T5 & 34500 & 21.2 \\
\hline
\end{tabular}
\\ $J_{\rm crit, 0}$ is the value of $J_{\rm crit}$ when $J_{\rm X, 21} = 0$.
\end{table}

The relationship between $J_{\rm X, 21}$ and $J_{\rm crit}$ that we recover is qualitatively very similar to
that found by \citet{it15} using a similar setup.\footnote{We carry out a more quantitative comparison in
Section~\ref{sec:prev} below.} This is no surprise, as the key features are easy to understand
physically. Collapsing gas that has reached a density of $10^{4} \: {\rm cm^{-3}}$ without forming enough 
H$_{2}$ to cool significantly will remain warm for the rest of the collapse, as above this density collisional
dissociation of H$_{2}$ becomes much more effective, while H$_{2}$ cooling becomes substantially less
effective \citep{bl03,io11}. Therefore, when $J_{21}$ is close to $J_{\rm crit}$, it is the amount of H$_{2}$ in gas
of around this density that determines whether or not the gas can successfully cool. This depends on the
formation rate of H$_2$, and since most of the H$_2$ forms via the associative detachment reaction
\begin{equation}
{\rm H^{-} + H \rightarrow H_{2} + e^{-}},
\end{equation}
this in turn depends on the formation rate of H$^{-}$. This ion forms via the radiative association reaction
\begin{equation}
{\rm H + e^{-} \rightarrow H^{-} + \gamma}
\end{equation}
and so the formation rates of H$^{-}$ and H$_{2}$ both depend directly on the fractional ionization of the 
gas. When $J_{\rm X, 21}$ is small, X-rays only provide a small amount of additional ionization, and so
the H$_{2}$ formation rate is barely affected. When $J_{\rm X, 21}$ is large, on the other hand, 
X-ray ionization makes a significant contribution to the ionization level of the gas. In the limit where 
this process dominates, the equilibrium fractional ionization is set primarily by the balance between
the X-ray ionization rate $R_{\rm X}$ and the radiative recombination rate $R_{\rm rec}$. Since
$R_{\rm X} \propto J_{\rm X, 21}$ and $R_{\rm rec} \propto x^{2}$, where $x \equiv n_{\rm e} / n$
is the fractional ionization of the gas, the result is that $x \propto J_{\rm X, 21}^{1/2}$. Therefore, in
this limit the H$_{2}$ formation rate $R_{\rm H_{2}} \propto J_{\rm X, 21}^{1/2}$, and so the radiation
field strength required to suppress H$_{2}$ cooling also scales as $J_{\rm X, 21}^{1/2}$
\citep[see also the similar analysis in][]{it15}.

\section{A reduced chemical network for modelling the influence of X-rays}
\label{sec:reduce}
Having established how $J_{\rm crit}$ varies as a function of $J_{\rm X, 21}$ in our fiducial model, we next 
analyze the full set of chemical reactions taking place during the evolution of the gas in each run using the
reduction technique described in Section~\ref{reduct}. For each run, we consider $\sim 18000$ different output times,
and use the results of our one-zone model to determine the weight of each reaction at each output time.
For each reaction, we then scan through this set of weights and find the maximum value for that reaction.
We repeat this procedure for each different value of $J_{\rm X, 21}$ and for runs with $J_{21} / J_{\rm crit} 
= 0.3, 1$ and 3. Finally, we construct our reduced network using only those reactions whose maximum weights 
exceed $\epsilon = 10^{-4}$ in at least one run. 

The reactions making up our final reduced network are shown in Table~\ref{tab:react}. The first point to note is that
reactions 1--26  are the same as those appearing in the reduced network we obtained in \citet{Glover15a}, where 
we considered the evolution of the gas in the absence of X-rays. This is unsurprising: the evolution of the gas when
$J_{\rm X, 21}$ is very small is essentially the same as in the case $J_{\rm X, 21} = 0$, and so it is natural that the
same chemical reactions will be important in this regime. 

\begin{table}
\caption{List of reactions with maximum reaction weights greater than $10^{-4}$ in at least one simulation \label{tab:react}}
\begin{tabular}{clcl}
\hline
No.\ & \multicolumn{3}{l}{Reaction} \\
\hline
1 & ${\rm H_{2} + \gamma}$ & $\rightarrow$ & ${\rm H + H}$ \\
2 & ${\rm H_2 + H}$ & $\rightarrow$ & ${\rm H + H + H}$ \\
3 & ${\rm H^{-} + H}$ & $\rightarrow$ & ${\rm H_{2} + e^{-}}$ \\
4 & ${\rm H_{2}^{+} + H}$ & $\rightarrow$ & ${\rm H_{2} + H^{+}}$ \\
5 & ${\rm H^{+} + e^{-}}$ & $\rightarrow$&$ {\rm H + \gamma}$ \\
6 & ${\rm H + e^{-}}$ & $\rightarrow$&$ {\rm H^{-} + \gamma}$ \\
7 & ${\rm H^{-} + \gamma}$ & $\rightarrow$&$ {\rm H + e^{-}}$ \\
8 & ${\rm H + H^{+}}$ & $\rightarrow$&$ {\rm H_{2}^{+} + \gamma}$ \\
9 & ${\rm H_{2}^{+} + \gamma}$ & $\rightarrow$&$ {\rm H^{+} + H}$ \\
10 & ${\rm H + H}$ & $\rightarrow$&$ {\rm H^{+} + e^{-} + H}$ \\
11 & ${\rm H^{-} + H}$ & $\rightarrow$&$ {\rm H + H + e^-}$ \\
12 & ${\rm H + e^{-}}$ & $\rightarrow$&$ {\rm H^+ + e^- + e^-}$ \\
13 & ${\rm H_{2}^{+} + He}$ & $\rightarrow$&$ {\rm HeH^+ + H} $ \\
14 & ${\rm H + He}$ & $\rightarrow$&$ {\rm H^+ + e^- + He} $ \\
15 & ${\rm H_2 + H^+}$ & $\rightarrow$&$ {\rm H_2^+ + H} $ \\
16 & ${\rm H_2 + He}$  & $\rightarrow$&$ {\rm H + H + He} $ \\
17 & ${\rm HeH^{+} + H}$ & $\rightarrow$&$ {\rm H_2^+ + He}$ \\ 
18 & ${\rm H + H + H}$ & $\rightarrow$&$ {\rm H_2 + H } $ \\
19 & ${\rm H^{-} + He}$  & $\rightarrow$&$ {\rm H + He + e^-} $ \\
20 & ${\rm H_{2}^{+} + H}$ & $\rightarrow$&$ {\rm H + H^{+} + H}$ \\
21 & ${\rm He + H^{+}}$ & $\rightarrow$&$ {\rm HeH^{+} + \gamma}$ \\
22 & ${\rm H^{-} + {\rm H^{+}}}$ & $\rightarrow$&$ {\rm H + H}$ \\
23 & ${\rm H_{2}^{+} + e^{-}}$ & $\rightarrow$&$ {\rm H + H}$ \\
24 & ${\rm HeH^{+} + e^{-}}$ & $\rightarrow$&$ {\rm He + H}$ \\
25 & ${\rm H^{-} + H^{+}}$ & $\rightarrow$&$ {\rm H_{2}^{+} + e^{-}}$ \\
26 & ${\rm H^{-} + e^{-}}$ & $\rightarrow$&$ {\rm H + e^{-} + e^{-}}$ \\
\hline
{\bf 27 }& ${\rm \bf H + \gamma} $&$ {\bf \rightarrow}$ & ${\rm \bf H^{+} + e^{-}}$ \\
{\bf 28 }& ${\rm \bf He + \gamma} $&$ {\bf \rightarrow}$ & ${\rm \bf He^{+} + e^{-}}$ \\
{\bf 29}& ${\rm \bf He^{+} + e^{-}} $&${\bf \rightarrow}$ & ${\rm \bf He + \gamma}$ \\
{\bf 30}& ${\rm \bf He^{+}  + H} $&${\bf \rightarrow}$ & ${\rm \bf He + H^{+}}$ \\
{\bf 31}& ${\rm \bf He^{+} + H^{-}} $&${\bf \rightarrow}$ & ${\rm \bf He + H}$ \\
{\bf 32}& ${\rm \bf He^{+} + \gamma} $&${\bf \rightarrow}$ & ${\rm \bf He^{++} + e^{-}}$\\
{\bf 33}&${\rm \bf He^{++} + e^{-}} $&${\bf \rightarrow}$ & ${\rm \bf He^{+} + \gamma}$\\
{\bf 34}&${\rm \bf He^{+} + H} $&${\bf \rightarrow}$ & ${\rm \bf HeH^{+} + \gamma}$\\
{\bf 35}&${\rm \bf HeH^{+} + \gamma}$&${\bf \rightarrow}$ & ${\rm \bf  He + H^{+}}$\\
\hline
\end{tabular}
\\ Reactions 1--26 are the same as in \citet{Glover15a}. Reactions 27--35
(highlighted in bold) are new in this study.
\end{table}

The new reactions that need to be added to our reduced reaction network in order to account for the influence of the
X-rays, numbers 27--35, are highlighted in bold in the table. We discuss the role that these reactions play in the
chemistry below.

\begin{itemize}
\item[]{\noindent \bf X-ray photoionization of H and He (reactions 27, 28)} \\
The most important effect of the X-rays is of course the additional ionization that they provide to the gas. Although
neutral atomic hydrogen (H) is substantially more abundant in the collapsing gas than neutral atomic helium (He),
the much larger size of the He photoionization cross-section at X-ray energies means that the rate of direct
photoionization is similar for both species. However, the majority of the ionization is actually secondary ionization
resulting from the collision of the energetic photoelectrons with further H and He atoms, and in this case the 
difference in abundances between H and He results in most of the secondary ionizations being ionizations of H.
Ultimately, therefore, most of the free electrons that are produced and that catalyze H$_{2}$ formation come from
the ionization of hydrogen. Despite this, it is important to retain helium in the model, since without it we will
significantly underestimate the rate of secondary ionization. \\

\item[]{\noindent \bf He$^{+}$ recombination (reaction 29)} \\
If we include He photoionization, then naturally we must also include He$^{+}$ recombination in order to keep 
our chemical model consistent. However, its inclusion in our model prompts the question of which rate coefficient
we should adopt for this process: should this be the case A rate, the case B rate, or some intermediate treatment?
We examine this in detail in Section~\ref{sec:rec} below. \\

\item[]{\noindent \bf He$^{+}$ charge transfer with H (reaction 30)}  \\
When the fractional ionization of the gas is small, this reaction can become competitive with reaction 29 as a 
sink for He$^{+}$, despite the fact that the latter has a much larger rate coefficient. The inverse reaction
\begin{equation}
{\rm H^{+} + He \rightarrow H + He^{+}}
\end{equation}
is unimportant, however, as its high endothermicity means that its rate is strongly suppressed at the temperatures
found in the collapsing gas. \\

\item[]{\noindent \bf Mutual neutralization of He$^{+}$ with H$^{-}$ (reaction 31)} \\
The rate coefficient for this reaction is slightly larger than that for the corresponding reaction with H$^{+}$ (reaction 22),
and so in some circumstances this can be a non-negligible destruction mechanism for H$^{-}$ despite the fact that in
general, $n_{\rm H^{+}} \gg n_{\rm He^{+}}$. \\

\item[]{\noindent \bf He$^{+}$ photoionization and He$^{++}$ recombination (reactions 32, 33)} \\
Our reduction algorithm finds that X-ray photoionization of He$^{+}$ to He$^{++}$ is marginally important, despite the
low abundance of He$^{+}$ relative to He in most of our models. However, we note that taking a slightly larger value
of $\epsilon$ would serve to eliminate this reaction from our reduced network, and so it is possible that our method
is being slightly too conservative here. If this reaction is included, then of course it is also necessary to include
He$^{++}$ recombination, as otherwise the network will not be physically consistent and the He$^{++}$ abundance
will grow without limit.  \\

\item[]{\noindent \bf HeH$^{+}$ formation from He$^{+}$ and H (reaction 34)} \\
We found in our previous analysis  that in some circumstances, it was necessary to account for the formation 
and destruction of the HeH$^{+}$ ion. However, in the conditions studied in that analysis, the He$^{+}$ abundance was
always tiny and hence HeH$^{+}$ formation by the radiative association of He$^{+}$ with H was never important,
even though the rate coefficient for this reaction is $\sim 10^{4}$ times larger than for the corresponding reaction between
H$^{+}$ and He (reaction 21). However, once we account for X-ray photoionization, we recover far larger He$^{+}$
abundances, particularly when $J_{\rm X, 21}$ is large, and hence find that this reaction becomes important. \\

\item[]{\noindent \bf HeH$^{+}$ photodissociation (reaction 35)} \\
The increased importance of reaction 34 in the presence of X-rays leads to us producing substantially more HeH$^{+}$
than when X-rays are absent. This increases the overall importance of HeH$^{+}$ in the chemical network and therefore
means that we need to model its destruction more carefully. As a result, our reduction algorithm finds that when X-rays are
present, this reaction becomes a marginally important destruction mechanism for HeH$^{+}$, even though in the absence
of X-rays it falls outside of our reduced set. 
\end{itemize}

\section{Sources of uncertainty}
\label{sec:uncertain}
In constructing our one-zone model we have made a number of choices that may introduce uncertainties into our
estimates for $J_{\rm crit}$. The impact of some of these (e.g.\ the choice of rate coefficients in cases where the
true value is uncertain) has already been explored in \citet{Glover15b}. Here, we examine the effect of uncertainties
specifically related to our treatment of X-ray photoionization. Note that our focus is on uncertainties arising from how 
we construct our numerical model, rather than astrophysical uncertainties such as the value of $J_{\rm X, 21}$ itself.
 
\subsection{Shielding length}
\label{shield}
One important source of uncertainty in one-zone models of the direct collapse process is the treatment of H$_{2}$
self-shielding and X-ray absorption. Commonly, the required column densities of H$_{2}$, H, He, etc.\
are obtained in a one-zone model by multiplying the local volume densities of H$_{2}$, H, He, etc.\ by a suitably
chosen characteristic length scale $L_{\rm c}$. In our fiducial model, we take this to be the Jeans length of the gas at the current
density and temperature, i.e.\ $L_{\rm c} = \lambda_{\rm J}$; we note that \citet{latif15} make a similar assumption. 
However, other authors have made different choices for this length scale.
For example, \citet{io11} and \citet{it15} assume a characteristic scale length of $L_{\rm c} = \lambda_{\rm J} / 2$. 
More recently, \citet{wgh16} recommend an even smaller value,
$L_{\rm c} = \lambda_{\rm J} / 4$. In reality, all of these choices are at best crude approximations for the behaviour of gas in a
real protogalaxy \citep{wgh11,hartwig15}.

\begin{figure}
\includegraphics[width=3.2in]{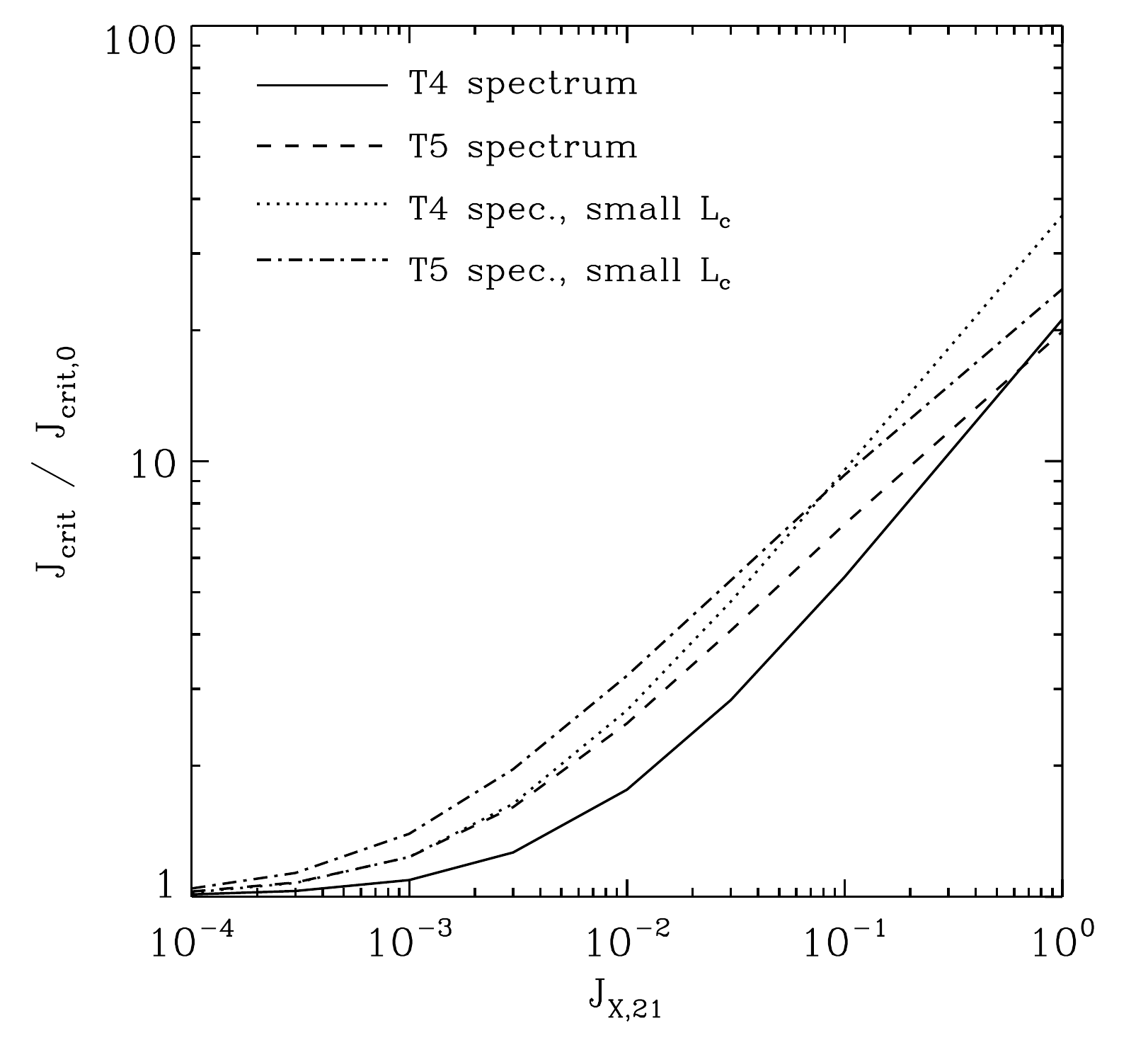}
\caption{Dependence of $J_{\rm crit}$ on $J_{\rm X, 21}$ in models where $L_{\rm c} = \lambda_{\rm J} / 2$,
computed for a T4 spectrum (dotted line) and a T5 spectrum (dot-dashed line). In both cases, the values of
$J_{\rm crit}$ are normalized by the value we obtain in the absence of X-rays given the same $L_{\rm c}$
and choice of UV-optical spectrum. For comparison, we also show the results from our fiducial model
for a T4 spectrum (solid line) and a T5 spectrum (dashed line).
\label{SMBH_Jcrit_lowL}}
\end{figure}

\begin{table}
\caption{Dependence of $J_{\rm crit}$ on $J_{\rm X, 21}$ in a model where $L_{\rm c} =  \lambda_{\rm J} / 2$.
\label{tab:jcrit_lowL}}
\begin{tabular}{rccc}
\hline
$J_{\rm X, 21}$ & Spectrum & $J_{\rm crit}$ & $J_{\rm crit} / J_{\rm crit, 0}$ \\
\hline
0.0 & T4 & 15.7 & 1.00 \\
$10^{-4}$ & T4 & 16.4 & 1.04 \\
$3 \times 10^{-4}$ & T4 & 17.8 & 1.13 \\
$10^{-3}$ & T4 & 21.9 & 1.39 \\
$3 \times 10^{-3}$ & T4 & 30.8 & 1.96 \\
0.01 & T4 & 50.5 & 3.22 \\
0.03 & T4 & 83.6 & 5.32 \\
0.1 & T4 & 146 & 9.30 \\
1.0 & T4 & 390 & 24.8 \\
\hline
0.0 & T5 & 851 & 1.00 \\
$10^{-4}$ & T5 & 870 & 1.02 \\
$3 \times 10^{-4}$ & T5 & 915 & 1.08 \\
$10^{-3}$ & T5 & 1050 & 1.23 \\
$3 \times 10^{-3}$ & T5 & 1390 & 1.63 \\
0.01 & T5 & 2280 & 2.68 \\
0.03 & T5 & 4050 & 4.76 \\
0.1 & T5 & 8130 & 9.55 \\
1.0 & T5 & 31200 & 36.7 \\
\hline
\end{tabular}
\\ $J_{\rm crit, 0}$ is the value of $J_{\rm crit}$ when $J_{\rm X, 21} = 0$.
\end{table}

To investigate the impact of our choice of characteristic length scale, we have re-run our models, keeping most parameters
the same but setting $L_{\rm c} = \lambda_{\rm J} / 2$. The results of these runs are plotted in Figure~\ref{SMBH_Jcrit_lowL}; 
we also include the results from our fiducial model, for reference. We see that in the runs with a smaller value of $L_{\rm c}$,
the impact of the X-rays is larger. In the runs with a T4 spectrum, decreasing $L_{\rm c}$ from $\lambda_{\rm J}$ to 
$\lambda_{\rm J} / 2$ leads to a 20--30\% increase in the value of $J_{\rm crit} / J_{\rm crit, 0}$ when $J_{\rm X, 21} \geq 0.01$.
In the runs with the T5 spectrum, the effect on $J_{\rm crit} / J_{\rm crit, 0}$ is significantly larger, between 50\% and 75\%
for the same range of $J_{\rm X, 21}$. However, if we examine the actual values of $J_{\rm crit}$ we obtain in these runs,
we find that in the runs with the T4 spectrum, they are only slightly larger than in our fiducial model, while in the runs with
the T5 spectrum, they are actually smaller (see Table~\ref{tab:jcrit_lowL}). In fact, most of the change in the behaviour of 
$J_{\rm crit} / J_{\rm crit, 0}$ is not due to any change in the value of $J_{\rm crit}$; instead, it is due to a decrease in the
value of $J_{\rm crit, 0}$ in the runs with a lower shielding length. This decrease is a result of the fact that in these runs,
we now have a smaller column density of H$_{2}$ shielding the collapsing gas, rendering H$_{2}$ self-shielding less effective.
When $J_{\rm X}$ is small, this effect dominates, and so $J_{\rm crit}$ is smaller than in the runs with larger $L_{\rm c}$.
When $J_{\rm X}$ is large, on the other hand, the reduced X-ray opacity and consequent increased ionization rate act to
offset this effect. 

\subsection{Values of $h\nu_{\rm min}$ and $\alpha$}
\label{res:xray}
The X-ray spectrum adopted in our simulations is specified by three numbers: the mean specific intensity at 1~keV,
$J_{\rm X, 21}$, the spectral slope $\alpha$, and the minimum energy cutoff $h\nu_{\rm min}$.\footnote{Technically,
it also depends on the maximum energy cutoff, $h\nu_{\rm max}$, but as previously noted our results are insensitive
to this choice provided that it is sufficiently large.} We have already examined how $J_{\rm crit}$ varies as we change
$J_{\rm X, 21}$. Here, we explore what happens if we vary the other two parameters.

In our fiducial model, $h \nu_{\rm min} = 0.5$~keV, but we have also carried out runs with $h \nu_{\rm min} = 0.1, 1$
and 2~keV. We find, in agreement with \citet{it15}, that our results are insensitive to our choice of $h\nu_{\rm min}$ 
provided that  $h \nu_{\rm min} \ll 1$~keV. This result is easy to understand.  In the density range relevant for determining
whether or not the collapsing gas will be able to form enough H$_{2}$ in order to cool, the inferred column density
of the protogalaxy ranges from a few times $10^{22} \: {\rm cm^{-2}}$ to a few times $10^{23} \: {\rm cm^{-2}}$. This
corresponds to an optical depth at 1~keV that ranges from $\tau \sim 1$ to $\tau \sim 10$, or a value at 0.5~keV
ranging from $\tau \sim 10$ to $\tau \sim 100$. Therefore, X-ray photons with energies around 1~keV contribute
significantly to the photoionization rate of the gas in the critical range of densities, but much softer X-ray photons do
not, as they are unable to penetrate deeply enough into the collapsing cloud. Consequently, the value we derive for
$J_{\rm crit}$ is insensitive to the presence or absence of these very soft X-ray photons, and hence is insensitive to
our choice of $h\nu_{\rm min}$, provided that it is less than 1~keV. 

\begin{figure}
\includegraphics[width=3.2in]{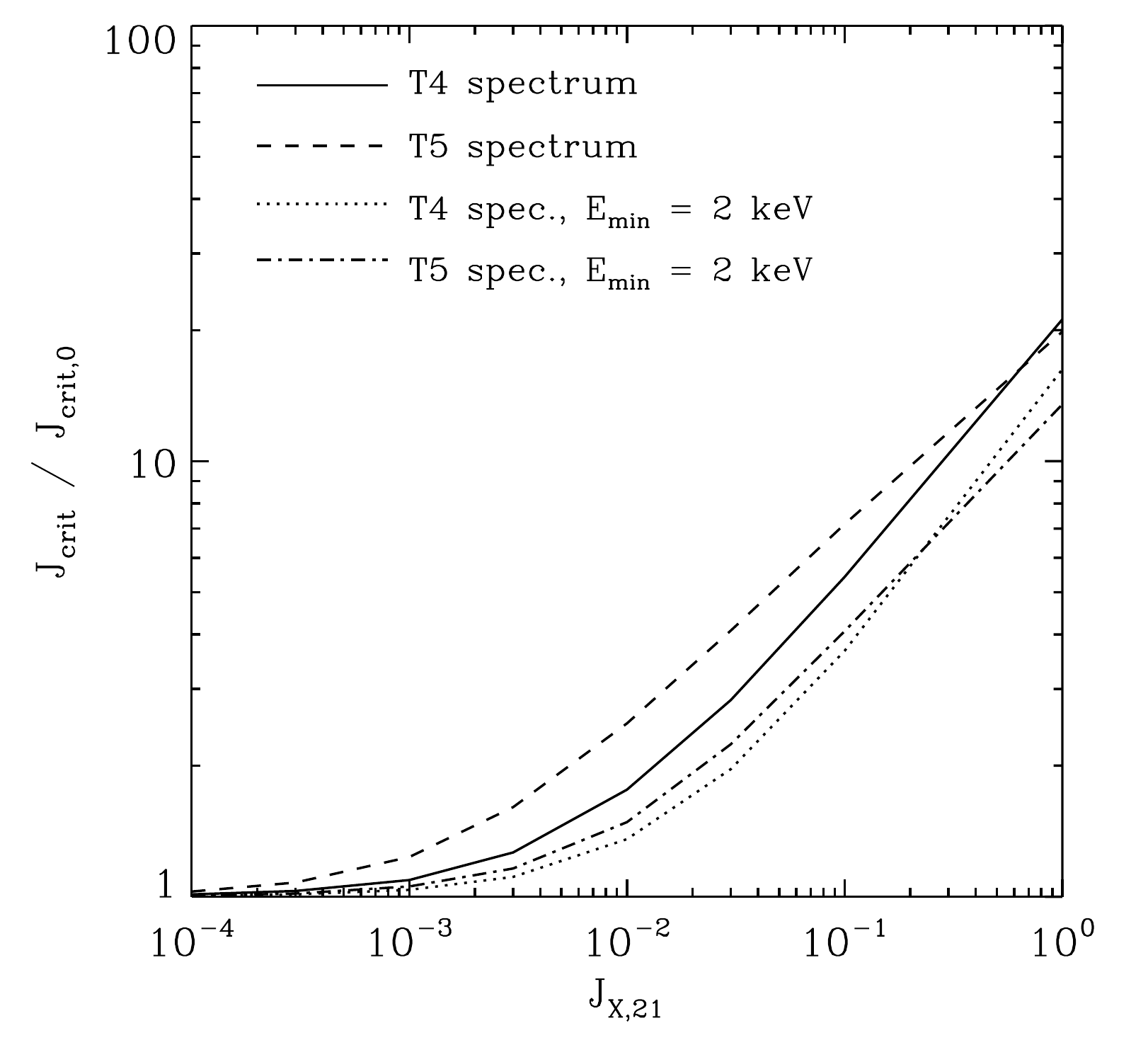}
\caption{Dependence of $J_{\rm crit}$ on $J_{\rm X, 21}$ in models where $h \nu_{\rm min} = 2$~keV,
computed for a T4 spectrum (dotted line) and a T5 spectrum (dot-dashed line). 
For comparison, we also show the results from our fiducial model for a T4 spectrum (solid line) and a 
T5 spectrum (dashed line). \label{SMBH_Jcrit_Emin}}
\end{figure}

If we increase the minimum energy further, to 2~keV, we do find a difference in the behaviour of $J_{\rm crit}$,
as illustrated in Figure~\ref{SMBH_Jcrit_Emin}. Increasing $h\nu_{\rm min}$ beyond 1~keV has the effect of reducing the X-ray
photoionization rate in the dense collapsing gas, since the 1--2 keV soft X-ray photons are responsible for
much of the ionization that occurs in our fiducial model. Consequently, we get a smaller fractional ionization
for a given $J_{\rm X}$, and recover a smaller value of $J_{\rm crit}$. The impact on $J_{\rm crit}$ varies
depending on $J_{\rm X}$ and our choice of spectrum, but typically, for X-ray backgrounds stronger
than $J_{\rm X, 21} \sim 0.01$, we obtain values of $J_{\rm crit}$ that are 30--70\% lower than in our fiducial
model.

\begin{figure}
\includegraphics[width=3.2in]{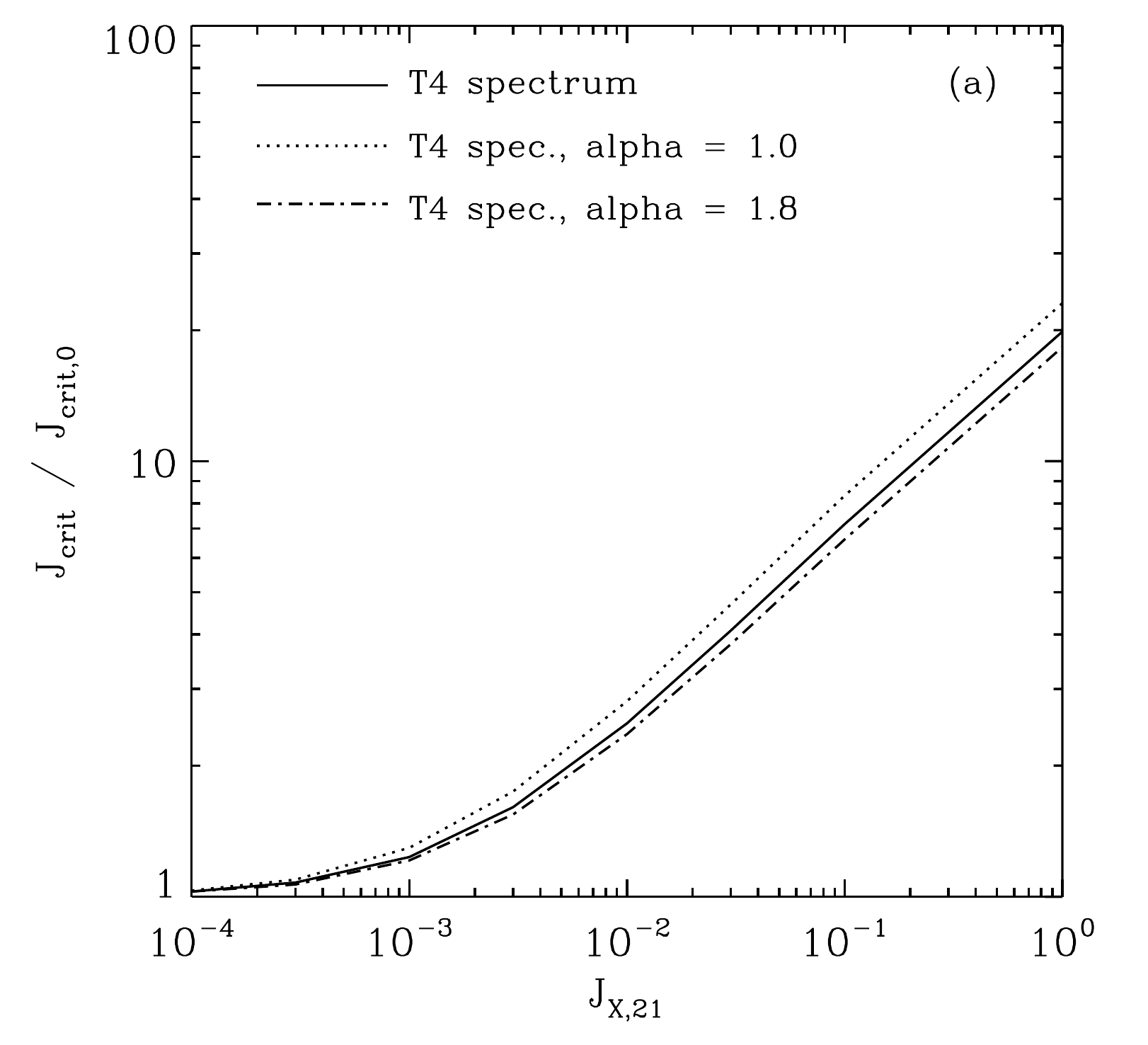}
\includegraphics[width=3.2in]{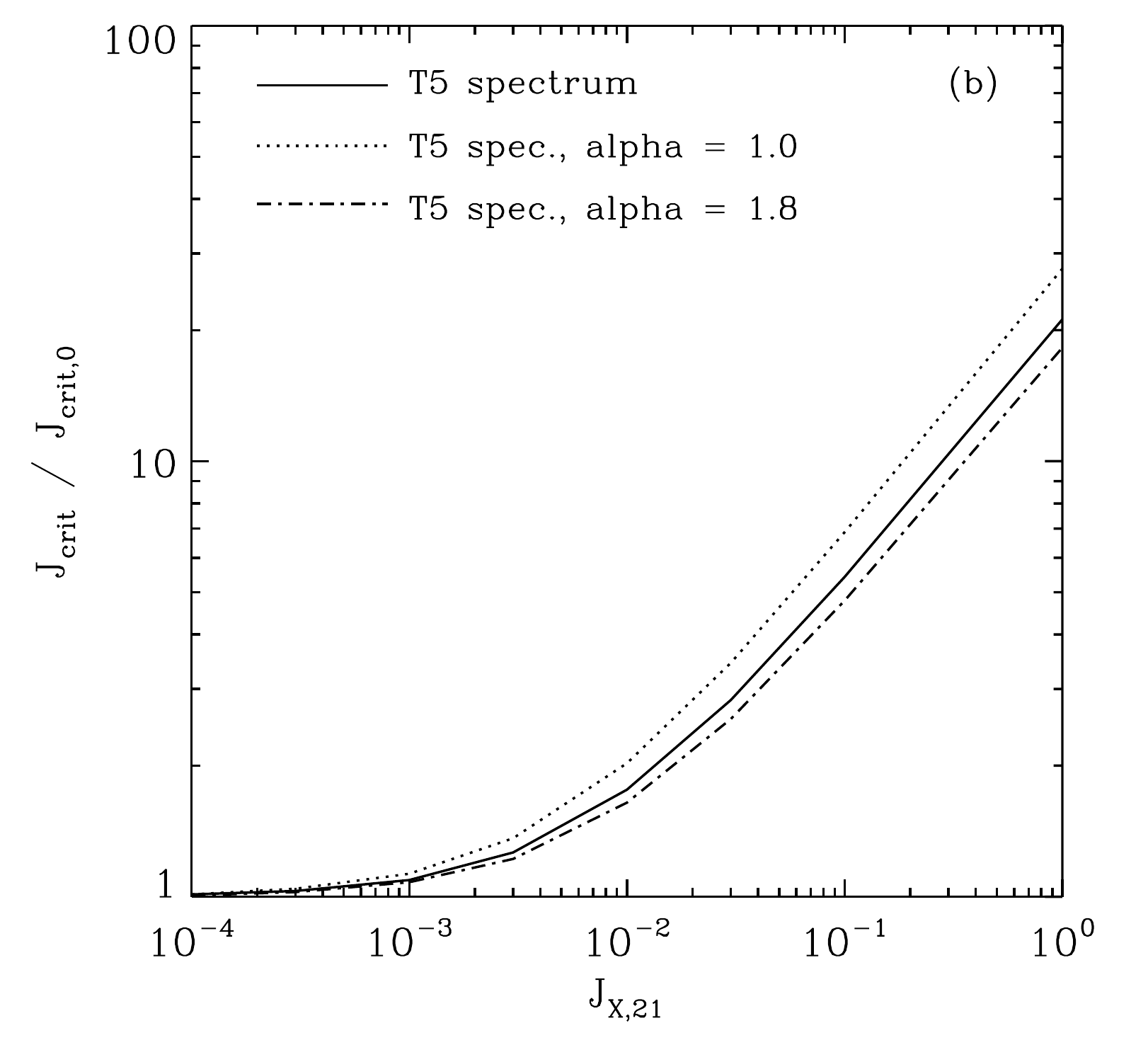}
\caption{(a) Dependence of $J_{\rm crit}$ on $J_{\rm X, 21}$ in models in which we vary $\alpha$,
the slope of the X-ray spectrum. We show results for $\alpha = 1.0$ (dotted line), $\alpha = 1.5$
(solid line; our fiducial model), and $\alpha = 1.8$ (dash-dotted line). All three runs were performed
using a T4 ultraviolet spectrum. 
(b) As (a), but for a T5 ultraviolet spectrum. \label{SMBH_Jcrit_alpha}}
\end{figure}

Varying the spectral slope while keeping the minimum X-ray photon energy fixed at $h \nu_{\rm min} = 0.5$~keV
also has an effect on $J_{\rm crit}$. In Figure~\ref{SMBH_Jcrit_alpha}a, we compare models run with $\alpha = 1.0$,
1.5 (our fiducial model), and 1.8, with a T4 UV spectrum. In  Figure~\ref{SMBH_Jcrit_alpha}b, we show a similar
comparison for the case of a T5 spectrum. In both cases, varying $\alpha$ has a very similar effect. Flattening the
X-ray spectrum by decreasing $\alpha$ leads to a mild increase in $J_{\rm crit}$ when $J_{\rm X, 21}$ is large.
Conversely, steepening the spectrum by increasing $\alpha$ causes $J_{\rm crit}$ to decrease. The reason for this
behaviour is that we normalize our spectra at 1~keV and as we have already seen, X-ray photons below 1~keV do
not penetrate deeply enough into the atomic cooling halo in order to significantly affect the chemistry of the gas
at the critical period during its collapse. Therefore, by decreasing $\alpha$, we increase the number of X-ray photons
at energies high enough to penetrate into and ionize the collapsing gas, and hence increase the ionization rate.
Similarly, by increasing $\alpha$, we decrease the number of ionizing photons reaching the gas at $n \sim 10^{3}
\: {\rm cm^{-3}}$ and hence decrease the ionization rate. 

Figure~\ref{SMBH_Jcrit_Emin} also demonstrates that the effect that varying $\alpha$ has on $J_{\rm crit}$ is
relatively modest. We find the largest impact for $J_{\rm X, 21} \sim 1$ for the T5 spectrum, but even here,
changing $\alpha$ from 1.0 to 1.8 changes $J_{\rm crit}$ by no more than 50\%. For lower values of $J_{\rm X, 21}$,
or for the T4 spectrum, we find a much smaller effect. We can therefore conclude that the uncertainty in the
spectral shape of the X-ray radiation field illuminating the atomic cooling halo introduces at most a 50\% uncertainty
into our determination of $J_{\rm crit}$. This is not insignificant, but is smaller than the uncertainty introduced by
our poor knowledge of the appropriate value of $J_{\rm X, 21}$ at high redshift.

\subsection{Choice of H and He cross-sections}
\label{sec:cross}
As mentioned in Section~\ref{sec:xray}, in our fiducial model we take values for the H and He photoionization
cross-sections from \citet{ost89} and \citet{ysd98,ysd01}, respectively. However, other authors \citep[e.g.][]{latif15}
have used instead the values given in the \citet{vf96} compilation of photoionization cross-sections. We have therefore
re-run our models using X-ray photoionization and photo-heating rates derived using these alternative cross-sections.
We find that in practice, this makes no significant difference to the results of our study. When $J_{\rm X}$ is large, we
obtain slightly larger values for $J_{\rm crit}$ if we use the \citet{vf96} cross-sections, but the difference is extremely
small, $\sim 0.2$\%. When $J_{\rm X}$ is small, the difference is likely even smaller, but as our iterative approach
only determines  $J_{\rm crit}$ to within 0.2\%, we cannot actually measure it reliably. In both cases, the uncertainty 
introduced is tiny compared with many of the other uncertainties in the model, and it is therefore safe to neglect it.

The reason that varying the H photoionization cross-section makes so little difference to the outcome becomes clear
if we compare the exact expression for $\sigma_{\rm H}$ given in \citet{ost89} 
with the analytical fit given in \citet{vf96}. The two expressions agree extremely well over the whole of the energy range
relevant for our study, differing by no more than about 0.2\%, and so regardless of which prescription we adopt, we
end up with almost exactly the same hydrogen photoionization rate. 

The case of helium, however, is a little more
interesting. The cross-section for He photoionization given by \citet{ysd98,ysd01} differs from that in \citet{vf96} by
around 10--20\% at the energies of interest. We might therefore expect the He photoionization rate to differ by a similar
amount, but in practice it does not; instead, the difference is only $\sim 1$\%. The reason for this is that although 
reducing the He photoionization cross-section decreases the probability of a given X-ray photon being absorbed
by He, it also decreases the optical depth of the gas, for the same reason. Therefore more X-ray photons are able
to reach the gas at  number densities $n \sim 10^{3}$--$10^{4} \: {\rm cm^{-3}}$, the behaviour of which determines 
whether or not direct collapse occurs. This increased number of photons largely offsets the decreased absorption
probability, and the result is that the He photoionization rate barely changes. Consequently, changing the He photoionization
cross-section from the \citet{ysd98,ysd01} prescription to the \citet{vf96} fit has almost no effect on the ionization state of the 
gas and hence no effect on the value of $J_{\rm crit}$.

Finally, it is important to note that the fact that our results are insensitive to small changes in $\sigma_{\rm H}$ and
$\sigma_{\rm He}$ does not imply that they are insensitive to large changes in these values. For example, if we use
the expression for $\sigma_{\rm He}$ given in \citet{ost89}, which is not intended for use at X-ray energies, then we
will dramatically overestimate the He photoionization rate and hence will also overestimate $J_{\rm crit}$ 
\citep[see e.g.\ the discussion in][]{latif15}.

\subsection{Secondary ionization model}
\label{sec:second}
In our fiducial model, we treat the effects of secondary ionization using values tabulated by \citet{dyl99} for a mix of H and He. 
However, many studies of the effects of X-rays at high redshift instead use the simple fitting functions given in \citet{ss85}, which
are based on calculations carried out using older atomic data and which assume that the X-ray photons have energies
$E \gg 100$~eV. We have therefore examined how sensitive our results are to this simplification.

We find that if we use the \citet{ss85} fitting functions in place of the \citet{dyl99} values, we recover values for $J_{\rm crit}$
that are between 3--5\% higher for X-ray field strengths $J_{\rm X, 21} \geq 0.01$. The shift towards higher values occurs
because the \citet{ss85} fitting functions predict that more of the X-ray energy goes into ionization and less into excitation of
H or He, or into heat. However, the difference with the \citet{dyl99} results is small for the photon energies relevant here, and
so the impact on $J_{\rm crit}$ is also small. 

\subsection{Treatment of He$^{+}$ recombination}
\label{sec:rec}
Recombination of an He$^{+}$ ion directly into its ground state produces a photon capable of ionizing either hydrogen or helium,
while recombination of the same ion into an excited state produces a photon capable of ionizing hydrogen (but not helium) with
a probability ranging from 66\% at high electron densities ($n_{\rm e} \gg 3 \times 10^{3} \: {\rm cm^{-3}}$) to 96\% at low 
electron densities ($n_{\rm e} \ll 3 \times 10^{3} \: {\rm cm^{-3}}$) \citep{ost89}. In our fiducial model, we account for these 
additional ionizing photons as part of our on-the-spot approximation, and also account for the fact that as some of the photons
produced by He$^{+}$ recombination into the ground state are absorbed by hydrogen, not helium, the effective He$^{+}$ 
recombination rate lies somewhere in between case A and case B. However, it is unclear whether this level of detail is
actually necessary in order to accurately determine $J_{\rm crit}$. We have therefore carried out additional runs in which we
set the He$^{+}$ recombination rate to either the case A or the case B rate and neglect any H ionizing photons produced by
He$^{+}$ recombination. (Note that we adopt the case B rate for H$^{+}$ recombination throughout, as we have already 
established that $J_{\rm crit}$ is highly sensitive to the choice of H$^{+}$ recombination rate coefficient -- see \citealt{Glover15b}).

We find that in practice, our results do not depend on how we treat He$^{+}$ recombination. We obtain extremely similar
values for $J_{\rm crit}$ regardless of whether we adopt the case A rate, the case B rate, or the more accurate treatment
described above. To help us understand why this is the case, we have examined the rates of the reactions responsible
for forming and destroying He$^{+}$ in the density range relevant for determining $J_{\rm crit}$, 
$n \sim 10^{3}$--$10^{4} \: {\rm cm^{-3}}$. We find that because of the low fractional ionization at these
densities, He$^{+}$ recombination is not the main process responsible for removing He$^{+}$ ions from the gas.
Instead, the removal of He$^{+}$ is dominated by charge transfer with atomic hydrogen, 
\begin{equation}
{\rm He^{+} + H} \rightarrow {\rm He + H^{+}}.
\end{equation}
Consequently, even fairly large changes in the He$^{+}$ recombination rate have only a very small influence on the
He$^{+}$ abundance, and hence on the contribution that He$^{+}$ makes towards the total fractional ionization. As
a result of this, changing the He$^{+}$ recombination rate has almost no influence on the H$_{2}$ formation rate, and
hence almost no influence on $J_{\rm crit}$.

\section{Comparison with previous studies}
\label{sec:prev}
The influence of X-rays on the value of $J_{\rm crit}$ has previously been investigated in studies by \citet{io11},
\citet{latif15} and \citet{it15}. However, as noted by \citet{latif15}, in their original study \citet{io11} adopted the approximate
helium photoionization cross-section given in \citet{ost89}, which is 1--2 orders of magnitude too large for photon energies of a
few keV. Therefore,
although their paper is important as it was the first one to point out the potential importance of X-ray ionization for
the direct collapse black hole scenario, their numerical results for the dependence of $J_{\rm crit}$ on $J_{\rm X}$
are incorrect.  For this reason, we do not compare our results with their study, but instead restrict our attention to
the more recent studies by \citet{latif15} and \citet{it15}. 

\citet{latif15} examine the effects of hard X-rays, with a minimum energy of 2~keV, within their 3D simulations of direct
collapse black hole formation. They find that even large X-ray fluxes have a relatively small impact on $J_{\rm crit}$,
increasing it by roughly a factor of 2 for $J_{\rm X, 21} = 0.1$ and a factor of 4 for $J_{\rm X, 21} = 1$. For comparison,
in our fiducial model, $J_{\rm crit} / J_{\rm crit, 0} \sim 5$ and $\sim 20$ for the same two values of the X-ray background field
strength. Some of the difference between the two models is due to the fact that they ignore X-ray photons at energies
below 2~keV, since these actually dominate the photoionization at the relevant densities. However, this does not completely
explain the difference: if we compare our results from runs with $h\nu_{\rm min} = 2$~keV (see Section~\ref{res:xray}) with 
their results, we see that our one-zone model still predicts roughly a factor of two larger impact on $J_{\rm crit}$ than they
find in their 3D runs. It is unclear why this difference exists, but it is possible that it reflects a systematic difference between
the effectiveness of X-rays in one-zone models compared to 3D models. In this context, it is interesting to note that 
\citet{it15} also examine the case where $h \nu_{\rm min} = 2$~keV and recover values for $J_{\rm crit} / J_{\rm crit, 0}$ 
at $J_{\rm X, 21} = 0.1$ and 1 that agree well with our results and disagree with the \citet{latif15} results.
Ultimately, additional 3D simulations of DCBH formation in the presence of X-rays will likely be needed in order to
understand the cause of this discrepancy.

\begin{figure}
\includegraphics[width=3.2in]{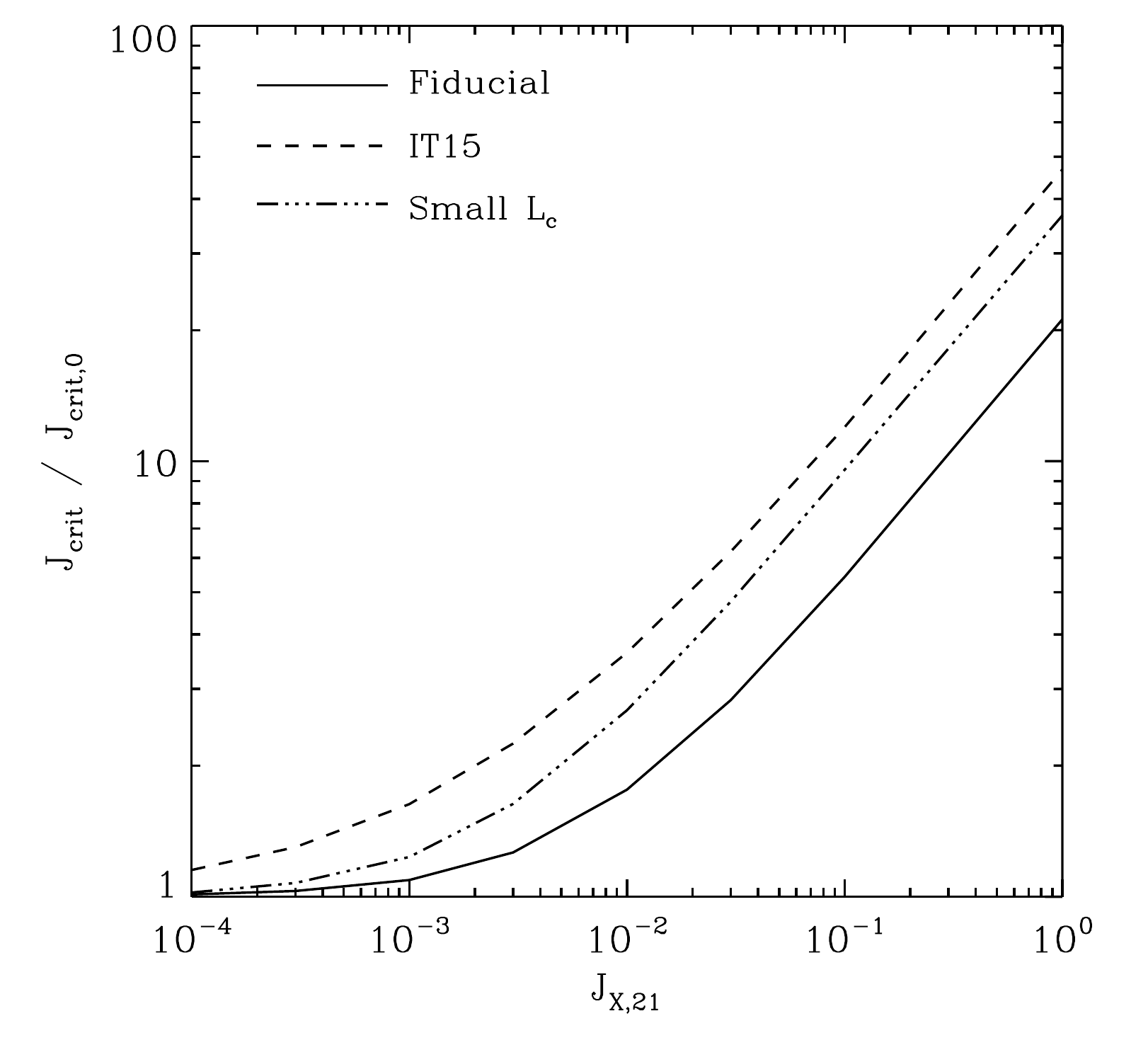}
\caption{$J_{\rm crit}$ as a function of $J_{\rm X, 21}$, plotted for our fiducial T5 model (solid line), the
\citet{it15} model with a T5 spectrum (dashed line) and a modification of our model in which we set the
shielding length to $L_{\rm c} = \lambda_{\rm J} / 2$ to better match the procedure used in \citeauthor{it15}'s
study. Note that the difference between \citeauthor{it15}'s results at very small $J_{\rm X}$ and our own
is due to the fact that their fitting function does not accurately represent their results for $J_{\rm X, 21}
\sim 10^{-4}$: in their one-zone models, they find that $J_{\rm crit}/J_{\rm crit, 0} = 1$ here, in good agreement
with our results. \label{SMBH_Jcrit_IT}}
\end{figure}

Turning to the study by \citet{it15}, we begin by noting that they do not consider the case of a T4 spectrum, so we cannot
compare our results with theirs for that case. However, they do carry out runs with a T5 spectrum (and also spectra with
$T_{\rm eff} = 2 \times 10^{4}$ and $3 \times 10^{4}$~K). In Figure~\ref{SMBH_Jcrit_IT}, we compare the results from our 
fiducial T5 model (solid line) to the values predicted by the fitting function given by \citet{it15} for the T5 case.\footnote{This function is
$J_{\rm crit} / J_{\rm crit, 0} = (1 + J_{\rm X, 21} / 0.0021)^{0.62}$.} We see that \citet{it15} find values of $J_{\rm crit} / J_{\rm crit, 0}$ 
that are roughly a factor of two larger than ours when $J_{\rm X}$ is large. There is also a small discrepancy when $J_{\rm X}$
is very small, but this simply reflects the fact that their fitting function overestimates $J_{\rm crit}$ slightly when $J_{\rm X, 21}
\ll 10^{-3}$ compared to their one-zone results \citep[see e.g. Figure 1 in][]{it15}.

Much of the difference between the results from \citet{it15} at large $J_{\rm X}$ and those from our fiducial model is due to the 
different assumptions we make regarding the shielding length. In our fiducial model, we assume that $L_{\rm c} = \lambda_{\rm J}$, 
whereas in their model, \citet{it15} assume that  $L_{\rm c} = \lambda_{\rm J} / 2$. As we have already seen in Section~\ref{shield}, this 
difference has a significant effect on the value of $J_{\rm crit}$. If we compare their results with those from a run in which we also
set $L_{\rm c} = \lambda_{\rm J} / 2$ (the dot-dashed line in Figure~\ref{SMBH_Jcrit_IT}), we find much better agreement, with a 
maximum difference of around 20\% when $J_{\rm X, 21} = 1$. This remaining difference is likely due to differences in the chemical
model used in the different simulations: as we have already seen in \citet{Glover15a,Glover15b}, differences in network design and
the choice of chemical rate coefficients can easily introduce uncertainties of this size or larger into $J_{\rm crit}$.

\section{Summary}
\label{sec:end}
In this paper, we have examined the impact of a soft X-ray background on the strength of the UV radiation field required
in order to suppressed H$_2$ formation in an atomic cooling halo, thereby allowing for direct collapse black hole
formation to occur. We confirm the earlier results of \citet{it15} that the X-rays have a significant effect on the value of
$J_{\rm crit}$ for X-ray field strengths $J_{\rm X, 21} > 0.01$ and that in the limit of large $J_{\rm X, 21}$, the critical
UV field strength scales as $J_{\rm crit} \propto J_{\rm X, 21}^{1/2}$. However, we recover values of $J_{\rm crit} /
J_{\rm crit, 0}$ (the value in the absence of an X-ray background) that are up to a factor of two smaller than those reported
by \citet{it15}. This difference is largely due to the different assumptions we make regarding the amount of shielding
against X-ray photoionization provided by the gas within the atomic cooling halo: we adopt a shielding length $L_{\rm c}
= \lambda_{\rm J}$, whereas \citet{it15} adopt a value a factor of two smaller, $L_{\rm c} = \lambda_{\rm J}/2$. This
sensitivity to the adopted shielding prescription emphasizes the need for the effect of X-rays to be re-examined in 3D
studies that can accurately account for the true 3D gas distribution, but this lies outside of the scope of the present study.

We have also used the reaction-based reduction technique developed by \citet{Wiebe03} and first applied to DCBH formation by
\citet{Glover15a} to identify the minimum subset of reactions needed to accurately determine $J_{\rm crit}$ from
simulations of the kind carried out here. This minimum subset consists of all 26 of the reactions in the reduced network
of \citet{Glover15a}, plus an additional 9 reactions that are required when $J_{\rm X} \gg 0$. 

Finally, we have carefully examined a number of possible sources of uncertainty in the modelling of the X-ray background
and its effect on the collapsing gas. We show that the results we recover for $J_{\rm crit}$ are independent of our choice
of parameterization for the H and He photoionization cross-sections, provided we choose fits that are appropriate for use
at X-ray energies. Our results are also independent of the accuracy of our treatment of He$^{+}$ recombination, as in
practice, charge transfer with atomic hydrogen is a more important loss route for He$^{+}$ in the physical conditions of 
interest. We also demonstrate that $J_{\rm crit}$ is only very weakly dependent on the prescription used to treat
secondary ionization. 

On the other hand, we demonstrate that our results are quite sensitive to the way in which shielding by neutral hydrogen
and helium elsewhere in the halo is accounted for, as already noted above. In addition, the values of $J_{\rm crit}$ we
recover depend on the slope assumed for the X-ray background, although varying the spectral index from $\alpha = 1$
to $\alpha = 1.8$ alters $J_{\rm crit}$ by at most 50\% in the limit of large $J_{\rm X}$. Finally, our results are independent
of the low energy cutoff chosen for the X-ray spectrum, $h\nu_{\rm min}$, provided this is less than 1~keV. However, increasing 
it above 1~keV leads to a substantial decrease in the values we obtain for $J_{\rm crit}$ when $J_{\rm X}$ is large.

\section*{Acknowledgements}
The author would like to thank B.\ Agarwal, J.\ Wolcott-Green, K.\ Inayoshi and J.\ Mackey for useful discussions on the topics of X-ray 
photoionization and DCBH formation. Financial support for this work was provided by the Deutsche Forschungsgemeinschaft  via
SFB 881, ``The Milky Way System'' (sub-projects B1, B2 and B8) and SPP 1573, ``Physics of the Interstellar Medium'' (grant number GL 668/2-1), 
and by the European Research Council under the European Community's Seventh Framework Programme (FP7/2007-2013) via the 
ERC Advanced Grant STARLIGHT (project number 339177).

\end{document}